# Analysis on numerical stability and convergence of RANS turbulence models from the perspective of coupling modes

Yilang Liu[1,2], Weiwei Zhang[2, *], Zhenhua Xia[1]

[1] *Department of Engineering Mechanics, Zhejiang University, Hangzhou 310027, China*

[2] *School of Aeronautics, Northwestern Polytechnical University, Xi'an 710072, China*

**Abstract:** Reynolds-averaged Navier-Stokes simulations are still the main method to study complex flows in engineering. However, traditional turbulence models cannot accurately predict flow fields with separations. In such situation, machine learning methods provide an effective way to build new data-driven turbulence closure models. Nevertheless, a bottleneck that the data-driven turbulence models encounter is how to ensure the stability and convergence of the RANS equations in posterior iterations. This paper studies the effects of different coupling modes on the convergence and stability between the RANS equations and turbulence models. Numerical results demonstrate that the frozen coupling mode, commonly used in machine learning turbulence models, may lead to divergence and instability in posterior iterations; while the mutual coupling mode can maintain good convergence and stability in the process of iterations. This research can provide a new perspective to the coupling mode for machine learning turbulence models with RANS equations in posterior iterations.

## 1. Introduction

Turbulence is considered as the last major unsolved problem in classical physics[1]. Numerical simulation methods play an increasingly important role in the research of turbulent problems. However, direct numerical simulation (DNS) and large eddy simulation (LES) are expensive in computational storage and cost for high Reynolds number flows. Therefore, Reynolds-Averaged Navier-Stokes (RANS) simulation is still the dominant tool for industrial problems in the near future.

\* Corresponding author: aeroelastic@nwpu.edu.cn (Zhang Weiwei)

Traditional RANS models, especially based on the eddy viscosity models (SA model[2], k-ε/k-ω model[3]~[5], Menter's SST model[6]), have been widely used to solve turbulence problems in engineering. These models can generally predict reliable aerodynamic coefficients for attached flows around complex configurations. However, the key difficulty that traditional RANS models encounter is that they cannot accurately simulate turbulent flows with separations.

Consequently, in recent years, researchers have developed many data-driven turbulence closures, including tensor based neural network (TBNN) models[8]~[9], machine learning augmented SA models[10]~[13], Reynolds stress discrepancy models[14]~[16] and physics informed neural network (PINN) models[17]~[18]. The models are trained by machine learning techniques based on DNS or LES data, which can significantly improve the predicted accuracy for separated flows compared with traditional RANS models. Although the recently developed models have made remarkable achievements in turbulence simulation, a serious problem is how to ensure the numerical stability and convergence when a machine learning turbulence model is coupled with the RANS equations in posterior iterations. The TBNN model, constructed by Ling et al.[8], was able to reach convergence only based on the converged solution of traditional RANS models, and the authors only pointed out that future studies would investigate the feasibility of iterative convergence which the TBNN model was coupled at every iteration. Thompson et al. [19] found that in the channel flows directly substituting Reynolds stresses from DNS databases into RANS equations led to mean velocity fields with large numerical errors, and the errors would increase with the Reynolds numbers. Poroseva et al.[20] also confirmed that unphysical solutions would occur when the fixed Reynolds stresses were directly inserted to the RANS equations. Wu et al.[21] demonstrated that treating Reynolds stresses as an explicit source term in the RANS equations may result in the ill-conditioned problem, and they also considered that the implicit treatment of the Reynolds stresses (decompose Reynolds stresses into the eddy viscosity part and the nonlinear part) could effectively alleviate the convergence problem. Cruz et al.[22] proposed to replace the Reynolds stress tensor with its divergence, the Reynolds force vector, as the machine learning target, which can more accurately reconstruct the average flow fields from DNS data. More recently, Brener et al.[23] elaborately explored different methods to solve the ill-conditioned problem of the RANS equations according to the plane channel, the square duct and the periodic hill geometries, and results indicated that the RANS equations were still ill conditioned in the whole range of cases even when the Reynolds stress tensor was treated implicitly. They also recommended that the Reynolds force vector with implicit treatment using optimized eddy viscosity could mitigate the error propagation to the mean velocity field in RANS equations. Guo et al.[24] considered that both the model error and the propagation error existed in RANS simulations, and the authors also mathematically derived the propagation error and claimed that it was an inevitable factor affecting the convergence of RANS simulations.

Mean velocity fields will produce large numerical errors when the Reynolds stresses extracted from DNS data with small errors are fixedly injected to the RANS equations, which is a key issue that the machine learning turbulence models are encountering. Many researchers proposed different strategies to circumvent the problem. Most existing work adopted the idea that inserting the fixed Reynolds stresses directly into the RANS equations, which did not consider the coupling effect between the Reynolds stresses and the RANS equations. The commonly used approach is to make the RANS solver firstly converge to the steady state with traditional turbulence models, and re-converge the simulation using the high-fidelity Reynolds stresses. However, this cannot ensure the iterative convergence if the simulation is started from uniform flow-fields. From a new perspective of the coupling mode between turbulence models and the RANS equations, this paper compares the convergence properties with different types of coupling modes, and we find out that the frozen coupling mode may result in numerical divergence and instability for the RANS simulation at high Reynolds numbers.

The remainder of this paper is organized as follows. Section 2 introduces the CFD governing equations and the methodology of RANS coupling modes. Section 3 presents the numerical examples to verify the proposed method, and finally the conclusions are drawn in Section 4.

## 2. Methodology

The steady-state RANS equations for incompressible flows can be written as:

$$\frac{\partial u_i}{\partial x_i} = 0$$

$$u_j \frac{\partial u_i}{\partial x_j} = -\frac{\partial p}{\partial x_i} + \nu \frac{\partial^2 u_i}{\partial x_j \partial x_i} - \frac{\partial R_{ij}}{\partial x_j} \quad (1)$$

where $u$ and $p$ are the mean flow velocity and pressure; $\nu$ is molecular viscosity; and $R_{ij}$ is the Reynolds stress tensor, which needs to be modeled. The term $R_{ij}$ can be decomposed into a linear part and a nonlinear part based on the eddy viscosity hypothesis, which is defined as:

$$R_{ij} = -2\nu_t S_{ij} + R_{ij}^{\perp} \quad (2)$$

where $\nu_t$ denotes the turbulent eddy viscosity; $S_{ij} = (\partial u_i / \partial x_j + \partial u_j / \partial x_i)/2$ is the velocity strain rate tensor; and $R_{ij}^{\perp}$ is the nonlinear part of the Reynolds stresses. The momentum equation can be rewritten as:

$$u_j \frac{\partial u_i}{\partial x_j} = -\frac{\partial p}{\partial x_i} + \frac{\partial}{\partial x_j}\left[(\nu+\nu_t)\frac{\partial u_i}{\partial x_j}\right] - \frac{\partial R_{ij}^{\perp}}{\partial x_j} \quad (3)$$

The main purpose to construct turbulence models is to resolve the RANS equations by modeling the unresolved term $\nu_t$ or $R_{ij}$, which are injected into RANS equations in order to solve the mean velocity field. There are mainly two parts to build machine learning turbulence models: the priori model training and the posterior model prediction. The training step is to obtain high-fidelity Reynolds stresses according to DNS/LES data and construct the mapping between mean flow features and Reynolds stresses. The prediction step is to inject the turbulence model into the RANS equations and modify the Reynolds stresses in order to obtain more accurate mean flow-field. There are two types of coupling modes in the posterior step: frozen coupling mode and mutual coupling mode. Fig. 1 demonstrates the schematic for different coupling modes. The frozen coupling mode denotes the trained Reynolds stresses are fixedly inserted into the RANS equations, and there is no information exchange between turbulence models and the RANS equations. On the other hand, the mutual coupling mode means that turbulence models and the RANS equations are coupled at every iteration step from the initial uniform flow to the converged state.

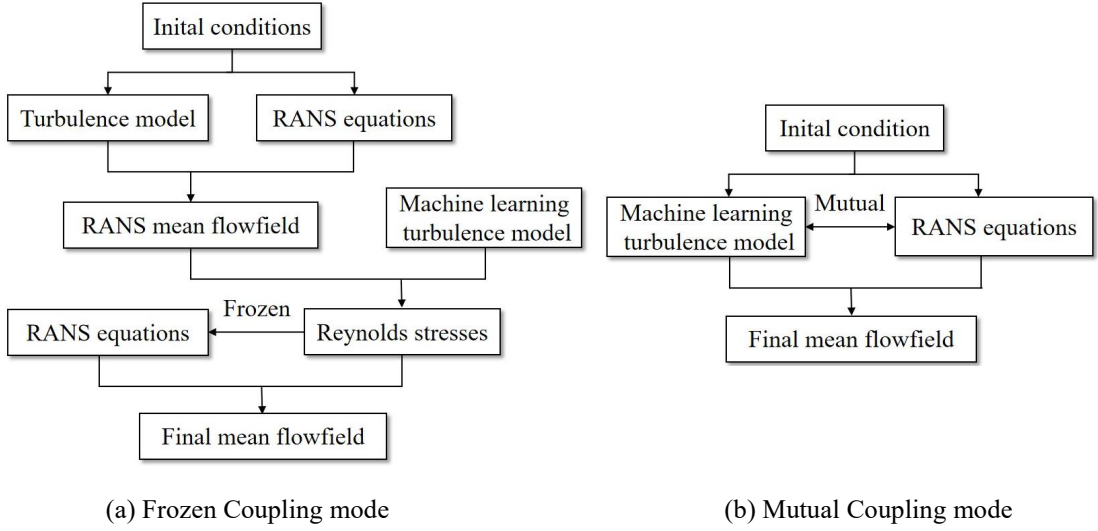

(a) Frozen Coupling mode    (b) Mutual Coupling mode

Fig. 1 Coupling modes between the machine learning turbulence model and the RANS equations

We discover that the mutual coupling mode can ensure the numerical convergence and the stability in the process of iteration from uniform flows, while the frozen coupling mode may induce divergence and instability in the simulation of separated flows at high Reynolds numbers. This paper mainly investigates the influence of two coupling modes on numerical convergence and stability when the RANS solver is initiated from uniform flows.

According to the Boussinesq hypothesis of linear eddy viscosity, the nonlinear part of the Reynolds stresses is ignored, and Equation (3) can be simplified as first-order closure models:

$$u_j \frac{\partial u_i}{\partial x_j} = -\frac{\partial p}{\partial x_i} + \frac{\partial}{\partial x_j}\left[(v+v_t)\frac{\partial u_i}{\partial x_j}\right] + \varepsilon_M \qquad (4)$$

where $\varepsilon_M$ denotes the modeling error introduced by the approximation of turbulence models. In the process of solving RANS equations by numerical methods, the propagation error $\varepsilon_P$ will also occur, which can be expressed as follows:

$$\left(u_j \frac{\delta}{\delta x_j} - \frac{\delta}{\delta x_j}\left[(v+v_t)\frac{\delta}{\delta x_j}\right]\right)u_i = -\frac{\delta p}{\delta x_i} + \varepsilon_M + \varepsilon_P \qquad (5)$$

The modeling error $\varepsilon_M$ has no influence on the convergence and the stability since it is a statistic error if a turbulence model is determined. And the propagation error $\varepsilon_P$ is produced during solving the RANS equations. $\varepsilon_P$ approaches to zero if the solver converges; otherwise, the error will continue to increase with the iteration of the solver and eventually lead to numerical instability or divergence.

This paper mainly studies the influence on numerical stability and convergence of the propagation error $\varepsilon_P$ with different coupling modes, and the model error $\varepsilon_M$ is not taken into consideration currently. For the mutual coupling mode, the propagation error is denoted by $\varepsilon_P^{MC}$, and Equation (5) can be rewritten as:

$$\left( u_j \frac{\delta}{\delta x_j} - \frac{\delta}{\delta x_j} \left[ (v+v_t) \frac{\delta}{\delta x_j} \right] \right) u_i = -\frac{\delta p}{\delta x_i} + \varepsilon_P^{MC} \qquad (6)$$

For the frozen coupling mode, we can firstly solve the RANS equations and obtain the turbulent eddy viscosity in the convergent state. And then, re-compute the RANS solver by fixedly inserting the converged turbulent eddy viscosity. The propagation error is denoted by $\varepsilon_P^{FC}$, that is

$$\left( u_j \frac{\delta}{\delta x_j} - \frac{\delta}{\delta x_j} \left[ (v+v_t^*) \frac{\delta}{\delta x_j} \right] \right) u_i = -\frac{\delta p}{\delta x_i} + \varepsilon_P^{FC} \qquad (7)$$

The RANS calculations are conducted with our in-house code which is based on the cell-centered finite volume method. In this paper, Roe scheme[29] is used to evaluate convective fluxes, and interface values are reconstructed by the second-order least-squares approach. The traditional SA model is used to solve the turbulent eddy viscosity. For the semi-discretized form of the governing equations, the implicit Symmetric Gauss-Seidel scheme[30] is adopted to march in time integral, and the local time stepping and residual smoothing techniques are also employed to accelerate convergence. Our in-house RANS solver has been validated completely by simulating many complex configurations in engineering at high Reynolds numbers, and the details can refer to our previous work[31]~[33].

# 3. Numerical Results

## 3.1 Curved backward facing step

The first test case is the curved backward facing step, and the definition of the geometry is the same as Ref. [34]. The step height is H and the Mach number is set as 0.2, and the Reynolds number is 13,700 based on H and inflow velocity. The computational mesh near the step is displayed in Fig. 2, which is locally refined. The total number of elements is 24,849, and the mesh growth rate is set as 1.1 in the boundary layer with the first grid height $8\times 10^{-4}$. We solve the RANS equations with the SA model and compare the results computed by the frozen coupling mode and the mutual coupling mode respectively. The convergence histories for different methods are shown in Fig. 3, and both coupling modes can ensure the numerical stability and convergence. Fig. 4 shows comparisons of streamlines and velocity distributions in the converged state, and we also draw scatter plots of the velocity magnitude of different coupling modes in the same coordinate system. The results show that nearly all points are coincided with the line *y=x*, which demonstrates that both coupling modes can achieve consistent and stable solutions. Therefore, the frozen coupling mode can ensure convergence and stability for flows at low Reynolds numbers.

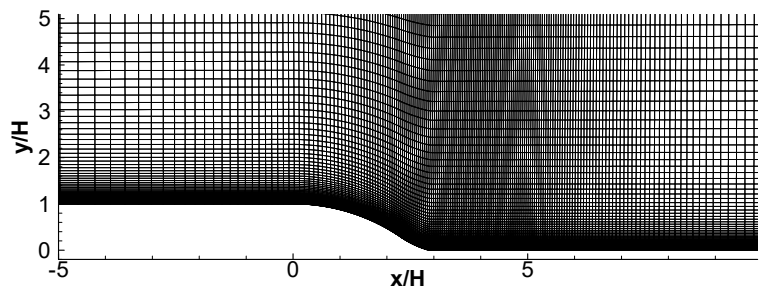

Fig. 2 Computational mesh near the step

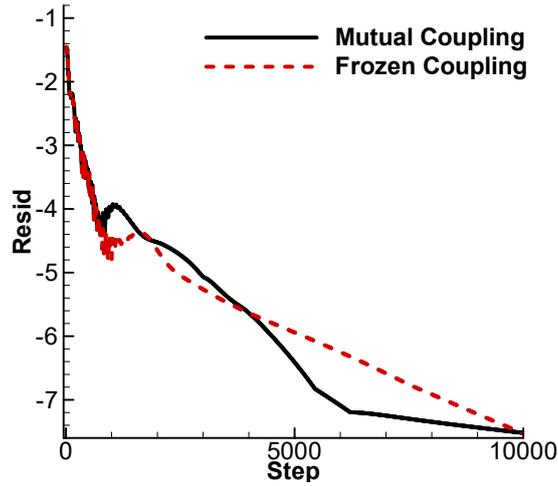

Fig. 3 Convergence of residuals with different coupling modes for curved backward facing step

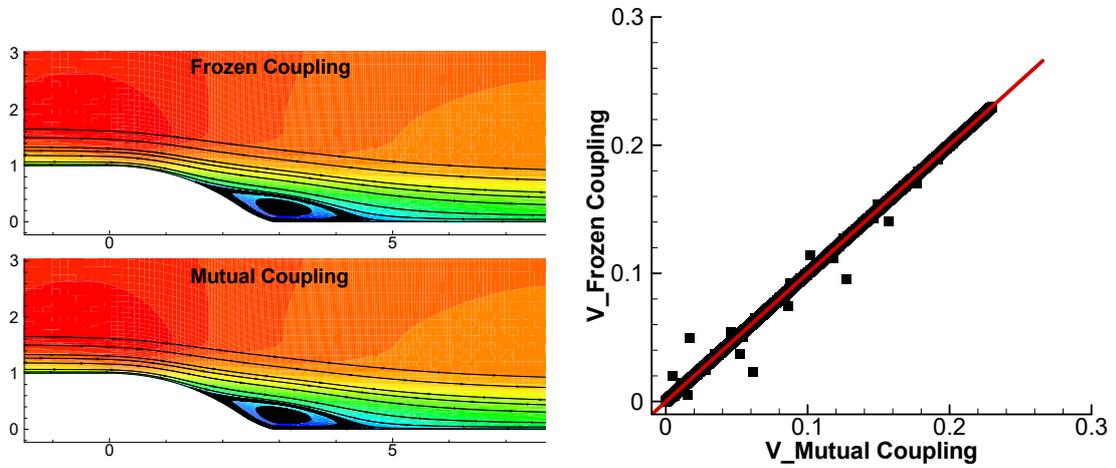

Fig. 4 Streamlines and scatter plots of velocity magnitude with different coupling modes for curved

backward facing step

### 3.2 Flow around a S809 airfoil at high Reynolds number

The second test case is the flow around a S809 airfoil at high Reynolds number. The computational mesh is shown in Fig. 5, in which the total number of elements is 36,077, with 400 nodes on the airfoil surface. There are 40 layers of mesh in the boundary layer with the growth rate 1.1, and the first grid height is $8 \times 10^{-6}$. The free stream Mach number is $Ma = 0.15$, and the Reynolds number is $Re = 2 \times 10^6$. We compare the results computed by the frozen coupling mode and the mutual coupling mode at different angles of attack.

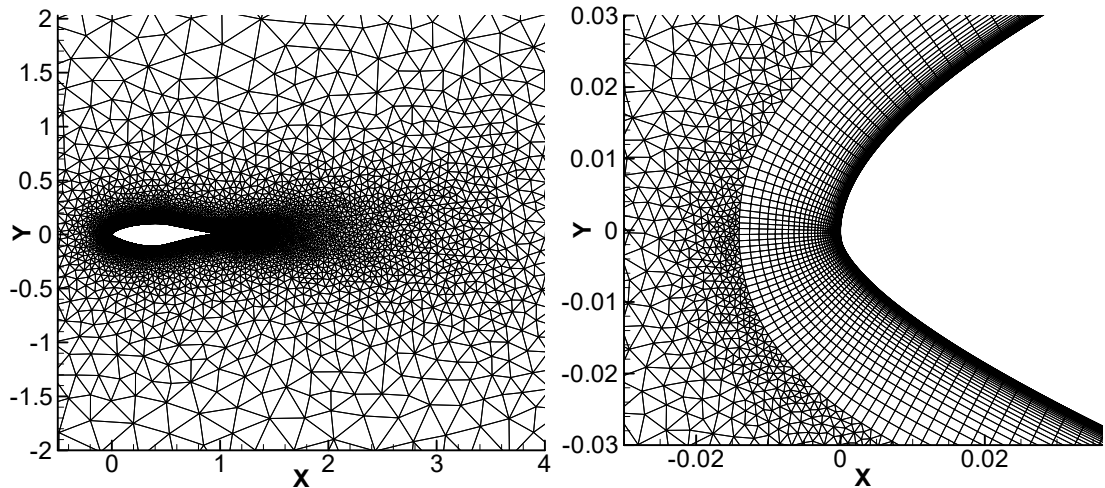

Fig. 5 Computational mesh for S809 airfoil

Firstly, for the angle of attack $\alpha = 8.2°$, the flow is attached and no separation exists in the whole field. The convergence of residuals and pressure coefficients with different coupling modes are depicted in Fig. 6 and Fig. 7. The results obtained by both coupling methods are almost coincided with each other, which indicates that the frozen coupling mode can ensure convergence and stability for attached flows, and it can achieve the same solutions compared with the mutual coupling mode.

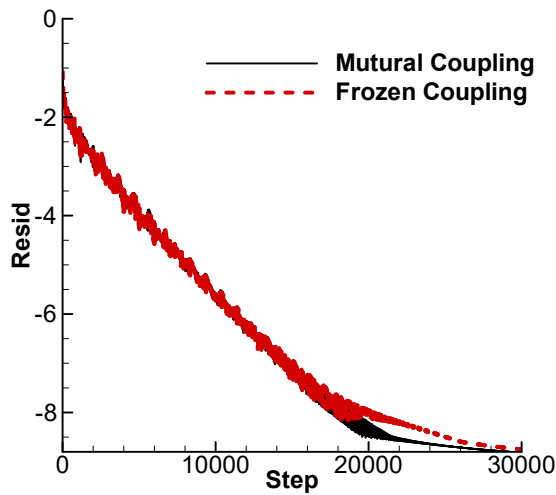 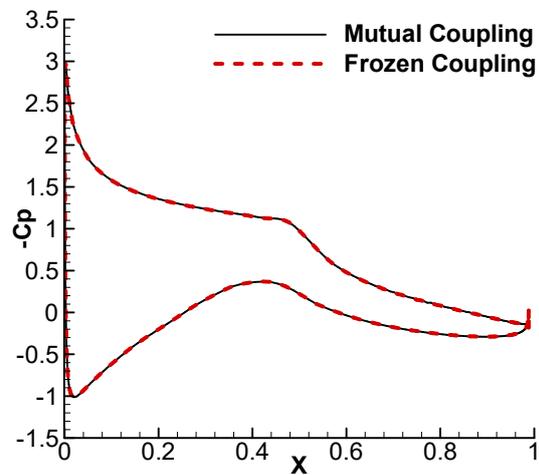

Fig. 6 Convergence of residuals with different coupling modes at 8.2° for S809 airfoil

Fig. 7 Pressure coefficients with different coupling modes at 8.2° for S809 airfoil

As the angle of attack increase to $\alpha = 14.2°$, a fixed vortex emerges on the upper surface at the trailing edge of the airfoil. We implement the RANS solver to 30,000 iterative steps by both coupling modes starting from uniform flows. Convergence histories and lift coefficients are displayed in Fig. 8 and Fig. 9, and comparisons of streamlines when the solver achieves the maximum iterative steps are shown in Fig. 10. The mutual coupling mode can ensure the numerical convergence and eventually reach the steady state with a small fixed separation region at the trailing edge. However, for the frozen coupling mode, both the residual and lift coefficients show sever oscillations and cannot achieve the steady state, and a wide range of detached unsteady vortex appears on the upper surface of the airfoil. The results indicate that the frozen coupling mode may encounter convergence problems for separated flows at high Reynolds numbers.

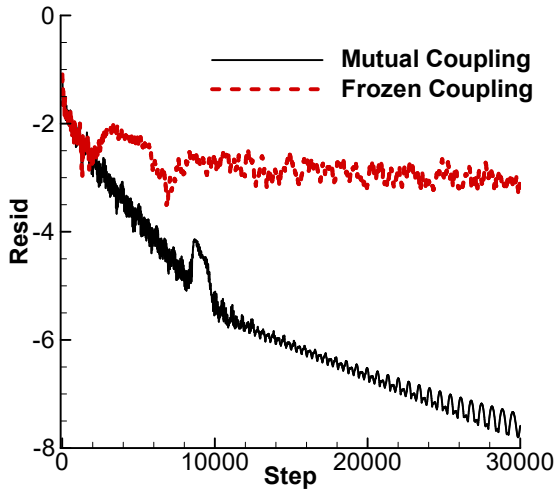
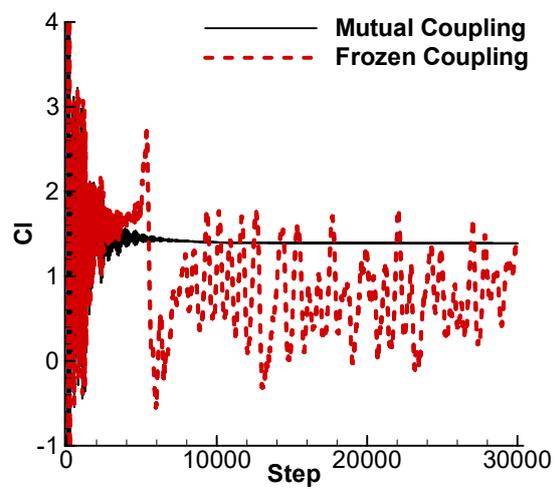

Fig. 8 Convergence histories for the S809 airfoil at $\alpha = 14.2°$ from uniform flows

Fig. 9 Response of lift coefficients for the S809 airfoil at $\alpha = 14.2°$ from uniform flows

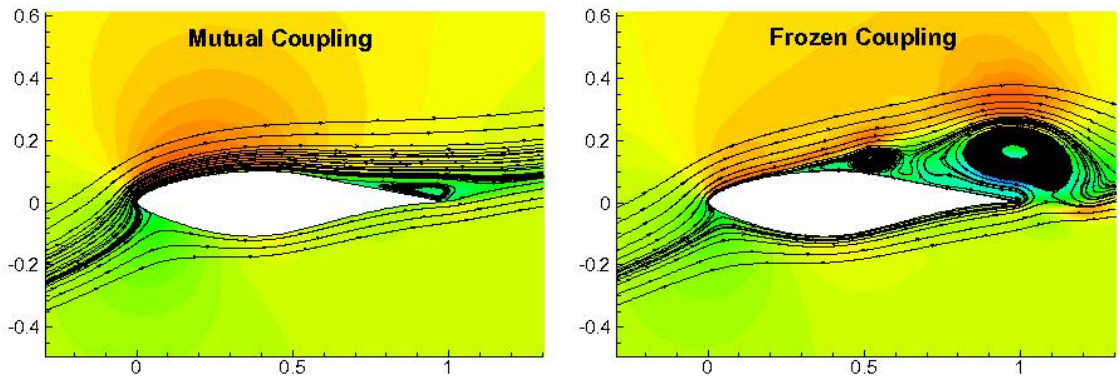

Fig. 10 Streamlines and velocity distributions for the S809 airfoil at $\alpha = 14.2°$ from uniform flows

We further investigate whether the frozen coupling mode can ensure the numerical stability when the RANS solver is based on the converged flow. Firstly, the mutual coupling mode is adopted to iterate 30,000 steps and the turbulent eddy viscosity is obtained in the convergent state. And then, the RANS solver is re-computed by the frozen coupling mode to a large number of iterative steps. Comparisons of convergence histories and lift coefficients for different coupling modes are shown in Fig. 11 and Fig. 12. With the increase of iterative steps, the mutual coupling mode can consistently maintain numerical stability and convergence. However, for the frozen coupling mode, the propagation error of the RANS equations will be gradually enlarged with the process of iterations, and both the residual and lift coefficients will eventually present numerical oscillations and non-convergence. Fig. 13 depicts the comparisons of streamlines near the trailing edge at the end of iterative step, which clearly shows that the vortex in the separated region is unsteady and constantly shaking. The results demonstrate that the frozen coupling mode may induce numerical instability even though the RANS solver is based on converged solutions.

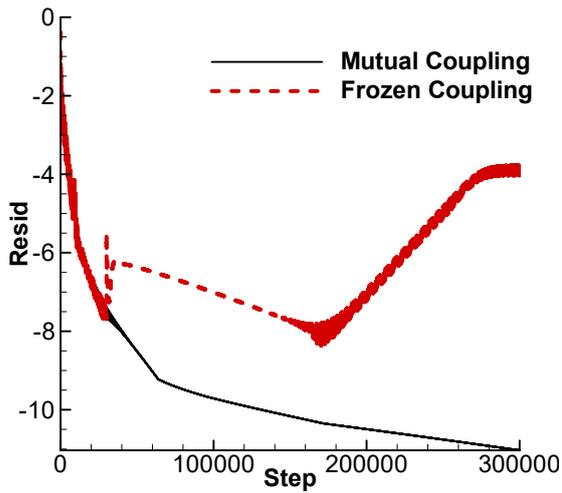
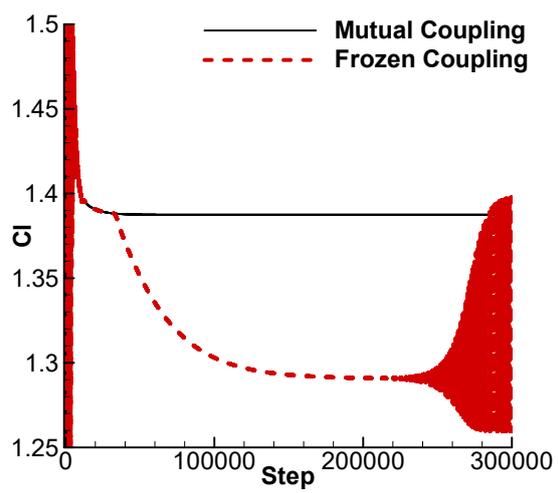

Fig. 11 Convergence histories for the S809 airfoil at $\alpha = 14.2°$ based on converged flows

Fig. 12 Response of lift coefficients for the S809 airfoil at $\alpha = 14.2°$ based on converged flows

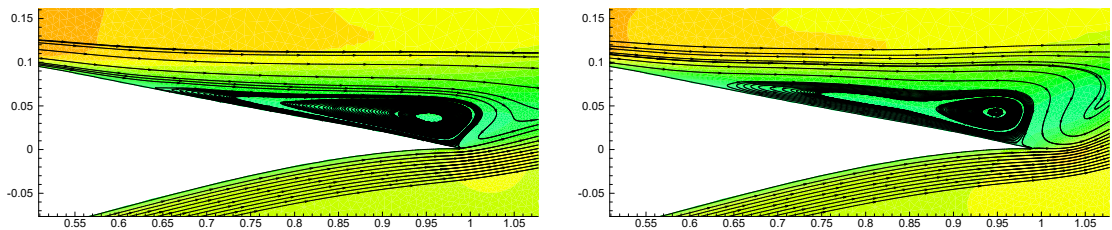

Fig. 13 Streamlines for the S809 airfoil at $\alpha = 14.2°$ based on converged flows. Left: mutual coupling mode; Right: frozen coupling mode.

As for the frozen coupling mode, the turbulent eddy viscosity is fixed, and there is no information feedback during the whole process of numerical calculation, which cannot eliminate propagation errors effectively. As the iterative step continues to increase, propagation errors may gradually accumulate and are enlarged, which will eventually lead to numerical non-convergence and instability. On the contrary, the mutual coupling mode can dynamically adjust the balance between the unclosed term (the eddy viscosity obtained by turbulence models), and the mean flow-field during iterative steps, so as to ensure the numerical stability and convergence of the RANS solution. For low Reynolds number flows, the molecular viscosity plays the dominated role and the turbulent fluctuation is relatively small; For the attached flows at high Reynolds numbers, the linear property of flow-field is dominant and the nonlinear property is relatively weak. Therefore, the frozen coupling mode can ensure numerical convergence to a certain extent in both cases. However, the separated flows at high Reynolds numbers have strong nonlinear characteristics, which are more sensitive to perturbation errors during the process of solution. Compared with the frozen coupling mode, the mutual coupling mode can ensure the numerical stability and convergence.

## 4. Conclusions

Reynolds stresses from DNS data are injected to the RANS solver as closure terms, which can cause significant errors for the mean velocity flow-fields. This paper analyzes this problem from a new perspective of the coupling modes. Based on the RANS equations with the traditional SA model, we compare the influences of the frozen coupling mode and the mutual coupling mode on numerical convergence and stability. As a result, the widely used frozen coupling mode may induce divergence and instability for separated flows at high Reynolds numbers even based on converged mean flow-fields; while the mutual coupling mode can dynamically adjust the balance between unclosed terms and mean flows, which can effectively eliminate propagation errors during iterative steps and ensure numerical convergence and stability. The curved back facing step and the high Reynolds flow around the airfoil are used to verify the proposed idea. In conclusion, this paper demonstrates the problem from a novel perspective that the RANS equations cannot obtain accurate mean flow-fields by injecting high fidelity Reynolds stresses. The proposed viewpoint can provide a new breakthrough to solve convergence and stability problems currently encountered in the machine learning modeling.